\documentclass[pra,twocolumn,showpacs,amsmath,amssymb,superscriptaddress,eqsecnum]{revtex4}
\usepackage{graphicx}
\usepackage{dcolumn}
\usepackage{bm,color}

\newcommand{\beq}{\begin{equation}}
\newcommand{\eeq}{\end{equation}}
\newcommand{\beqa}{\begin{eqnarray}}
\newcommand{\eeqan}{\end{eqnarray*}}
\newcommand{\beqan}{\begin{eqnarray*}}
\newcommand{\eeqa}{\end{eqnarray}}
\newcommand{\bra}[1]{\langle{#1}|}

\newcommand{\ket}[1]{|{#1}\rangle}

\newtheorem{theo}{Theorem}
\newcommand{\bqa}{\begin{eqnarray}}
\newcommand{\eqa}{\end{eqnarray}}
\newcommand{\nn}{\nonumber}
\newcommand{\nl}[1]{\nn \\ && {#1}\,}
\newcommand{\erf}[1]{Eq.~(\ref{#1})}

\newcommand{\du}{\partial}

\newcommand{\sq}[1]{[ {#1} ]}
\newcommand{\cu}[1]{\left\{ {#1} \right\}}

\newcommand{\an}[1]{\left\langle{#1}\right\rangle}

\newcommand{\tr}[1]{{\rm Tr}\sq{ {#1} }}
\newcommand{\s}[1]{\hat \sigma_{#1}}
\definecolor{nblue}{rgb}{0.2,0.2,0.7}
\definecolor{ngreen}{rgb}{0.2,0.6,0.2}
\definecolor{nred}{rgb}{0.7,0.2,0.2}
\definecolor{nblack}{rgb}{0,0,0}

\begin{document}
\title{Optimality of feedback control strategies for qubit purification}
\author{Howard M. Wiseman} 
\affiliation{Centre for Quantum Computer Technology, Australia}
\affiliation{Centre for Quantum Dynamics, Griffith
University, Brisbane, Queensland 4111 Australia}
\author{Luc Bouten} 
\affiliation{Physical Measurement and Control 266-33, California
Institute of Technology, Pasadena, CA 91125}
\date{\today}
\pacs{03.67.-a, 03.65.Ta, 02.50.Tt, 02.50.Ey}
\begin{abstract}
Recently two papers [K. Jacobs,  Phys. Rev. A {\bf 67}, 030301(R) (2003); 
H. M. Wiseman and J. F. Ralph, New J. Physics {\bf 8}, 90 (2006)]
 have derived control strategies
for rapid purification of qubits, optimized with respect to various goals. 
In the former paper the proof of optimality was not mathematically  
rigorous, while the latter gave only heuristic arguments for optimality.
In this paper we provide rigorous proofs of optimality in all cases, 
by applying simple concepts from optimal control theory, 
including Bellman equations and verification theorems.
\end{abstract}
\maketitle
\section{Introduction}
In the absence of decoherence, monitoring (that is, continuous-in-time weak 
measurement) of a qubit observable such as $\sigma_z$ 
will eventually purify the qubit.
However, the process of purification for finite times can be
affected by applying Hamiltonian controls. 
Based upon results for discrete (but non-projective)
 measurements, Jacobs  \cite{Jac03,Jac04} derived 
a strategy to {\em maximize the average purity} of the qubit
at any given time. His strategy requires the controller to maintain the qubit state with
$\an{\sigma_z}=0$.  Wiseman and Ralph \cite{WisRal06} considered two 
different goals: first to {\em maximize the fidelity} of the qubit with any given
state at any given time, and second, 
to {\em minimize the average time} required to attain a
given level of purity (or fidelity).  For the first goal, they argued that Jacobs' 
strategy (with a trivial modification) was again optimal.
For the second goal, they considered the  strategy 
which maintains $\an{\sigma_y} = \an{\sigma_z} = 0$, which 
is a no-control strategy provided that these conditions are met initially 
(and that any Hamiltonian not proportional to $\sigma_z$ can be nullified).
They showed that this outperforms Jacobs' control strategy, and argued
that it was the optimal strategy. 
However, none of these claims of optimality were proven using  
rigorous results from continuous time stochastic control theory.

Control strategies derived from heuristic arguments may well fail to 
attain optimality.  Some examples include continuous 
quantum error correction in Ref.~\cite{Ahn02} and gradual discrimination 
of non-orthogonal states in Ref.~\cite{Jac06} (in which Jacobs is well 
aware of the non-optimality). 
However, if a simple feedback strategy is conjectured to be 
optimal, then the truth of the conjecture can be demonstrated straightforwardly 
using standard techniques  from optimal control 
theory --- the so-called {\it verification theorems}.  The purpose of this 
paper is to highlight this fact, by showing that both of the above strategies 
(that of Jacobs, and that of Wiseman and Ralph),  
are indeed optimal for the different goals as claimed.
We introduce the system in Sec.~II, as well as reviewing some 
terminology and results from optimal control theory. In Sec.~III
we discuss and refine Jacob's problem, and prove the optimality of his 
strategy. In Sec.~IV we do likewise for Wiseman and Ralph's problem,
and we conclude with a brief discussion in Sec.~V.

\section{The system}
In continuous measurement theory a basic role is played by the quantum
filtering equation \cite{Bel88,BvH05,BGM04}, 
also known as a stochastic master equation \cite{WisMil93c,Wis96a}. 
This can be obtained 
from an underlying quantum Markov model describing the coupling of the 
system to an external environment. The filtering
equation is a stochastic differential equation whose solution at time $t$
is a (random) state matrix $\rho_t$, such that the quantity
$\tr{X\rho_t}$ corresponds to the conditional expectation of  an arbitrary operator 
 $X$ with
respect to the observations up to time $t$.  In a more
traditional quantum measurement perspective  (see for example Ref.~\cite{Wis96a}),
$\rho_t$ is known as the conditional system state, 
and represents the observer's beliefs about the system. 
The filtering equation also plays a fundamental role in 
quantum feedback control, because most control
problems separate into a filtering step and a 
classical control problem for the filtering 
equation \cite{BvH05, Bel88}. 
That is, the optimal feedback control algorithms  
make use only of $\rho_t$ and have thus 
been called state-based \cite{Doh00} and 
Bayesian \cite{WisManWan02} in the physics literature. 
In control theory such strategies are known as separated 
strategies.  
 
  The goal of the particular control problems
studied in \cite{Jac03,Jac04,WisRal06} is to maximize the rate of purification of the
conditional state $\rho_t$ in a suitable sense.
 We consider precisely the same system as in 
\cite{Jac03,Jac04,WisRal06}: monitoring one 
operator ($\s{z}$) for a qubit with the ability to control the 
Hamiltonian. In a solid-state context this monitoring could be realized by a quantum point 
contact \cite{Goa01}, and in an optical context it could arise 
from an off-resonant coupling of the qubit to the 
electromagnetic field which is observed continuously in time 
via homodyne detection \cite{HanStoMab05}. Here we denote by $\sigma_x,\sigma_y,\sigma_z$ 
the well-known Pauli matrices. We assume without loss of 
generality that the initial state $\rho_0$ satisfies 
$\tr{\rho_0\sigma_y}=0$. In that case we 
only need a control Hamiltonian proportional to $\sigma_y$. 
The quantum filtering equation for this model is 
\bqa
	d\rho_t &=& 
		-i{\Omega_t}[\sigma_y/2,\rho_t]\,dt +
		(\sigma_z\rho_t\sigma_z-\rho_t)\,dt  \nn \\ 
		&& +	(\sigma_z\rho_t+\rho_t\sigma_z-2\tr{\sigma_z\rho_t}
			\rho_t)\,dW_t. \label{QFE1}
\eqa
Here $\Omega_t$ is the control input and $W_t$ is the
innovations Wiener process \cite{Bel88,BGM04,BvH05}. 
It is related to the measurement record $Y_t$ on which $\rho_t$ is conditioned by 
\beq
dY_t =  2\tr{\rho_t \s{z}}dt + dW_t.
\eeq

Rewriting \erf{QFE1} in 
terms of the Bloch vector components $(x_t,y_t,z_t)={\rm 
Tr}[(\sigma_x,\sigma_y,\sigma_z)\rho_t]$, we obtain
  \bqa
  dz_t &=& 2 (1-z_t^2)\,dW_t - \Omega_t\,x_t\,dt, \\
  dx_t &=& -2x_t\,dt - 2z_tx_t\,dW_t + \Omega_t\,z_t\,dt,
  \eqa 
  and $y_t=0$. We now introduce polar coordinates 
$z_t = R_t\cos\theta_t$ and $x_t =R_t\sin\theta_t$.
It turns out to be convenient in calculations to work with 
$S_t = 1 -R_t^2$ instead of $R_t$. $S_t$ is sometimes known 
as the linear entropy of the qubit.
In these variables the filtering equation reads 
\bqa
   dS_t &=& -4S_t\left\{\left[1-(1-S_t)u_t^2\right]dt +\sqrt{1-S_t}\, u_t\, dW_t\right\}, \nn \\
&&    \label{eq system} \label{dS}  \\
   d\theta_t &=& \Omega_t  dt +f(\theta_t,S_t) dt + g(\theta_t,S_t) dW_t. \label{eqtheta}
\eqa
Here $f$ and $g$ are 
functions whose explicit form we will not need and we 
have written $u_t$ for  $\cos\theta_t$. 

The equations (\ref{eq system}) and (\ref{eqtheta}) are 
already too complicated to be amenable to optimal control.
In order to proceed, the papers \cite{Jac03,Jac04,WisRal06} assume that, 
rather than controlling $\Omega_t$, one can directly control $u_t$ at
every time to take any value in the interval $[-1,1]$.  This could be seen
as a singular limit of the above system, in the sense that an infinitely
strong driving field $\Omega_t$ would ostensibly allow one to 
set the angle $\theta_t$ to any desired value. This type of assumption is
commonly used to reduce open-loop optimal control problems in NMR to
explicitly computable form \cite{Navin}.  In the present case,
it is less clear that this is a reasonable approximation, as 
singular controls are incompatible with the assumptions used to
derive the quantum filtering equation \cite{Bel88,BvH05}.  
Our goal here, however, is to demonstrate optimality of the 
problems ultimately considered in \cite{Jac03,Jac04,WisRal06}, 
so we will adopt their assumptions in the following.

Before we proceed to the problems considered in 
\cite{Jac03,Jac04,WisRal06}, we review some 
terminology.  A control strategy of the form $u_t=u(t,S_t)$, where 
$u(t,s)$ is a deterministic function, is called a \emph{Markov} control. 
It is easily verified using the It\^o rule that for any function 
$f(t,s)$ that is $C^1$ in $t$ and $C^2$ in $s$, we have
\bqa
	\mathbf{E}_s \bigl(f(t,S_t)\bigr) &=& 
	\mathbf{E}_s  \int_0^t\!\left[
		\frac{\partial f}{\partial t'}(t',S_{t'})+
		\mathcal{L}^{u_{t'}}f(t',S_{t'})
	\right]\!dt' \nl{+} f(0,s) ,
\eqa
which is known as {\it Dynkin's formula}. 
Here $\mathbf{E}_s(X)$ denotes ${\bf E}(X|S_0=s)$, the expectation of $X$ 
given the initial condition $S_0 =s$.
Moreover, for any $v\in[-1,1]$, we have defined 
the differential operator ${\cal L}^v$ by 
\beq
	\mathcal{L}^vf(s) = -4s\frac{\partial f(s)}{\partial s} +  v^2{\cal D}f(s),
\eeq
where
\beq{\cal D}f(s) = 
		8s^2(1-s)\frac{\partial^2f(s)}{\partial s^2}
		+4s(1-s) \frac{\partial f(s)}{\partial s}.
\eeq

\section{Jacobs' problem}
Jacobs \cite{Jac03,Jac04} considered the following 
control problem: Fix an arbitrary terminal time 
$T>0$, and find a feedback control $u_t$ (with 
values in $[-1,1]$) that maximizes the purity of the qubit at time $T$. 
Specifically, Jacobs chose a goal of minimizing the cost function ${\bf E}_s(S_T)$, 
the expected value of the random variable $S_T$, for $S_0=s$. This choice of cost
function was made ``to keep the calculations tractable'' \cite{Jac03}. While this choice 
gives a well-defined control problem with a well-behaved solution (as discussed below), it 
was pointed out by Wiseman and Ralph \cite{WisRal06} that the 
cost function ${\bf E}(S_T)$ lacks a clear physical motivation. The 
reason is that it is a nonlinear function of the 
quantum state, and hence does not correspond to the expected outcome 
of any physical process that can be performed on the qubit. 

Say the qubit is being purified for some particular purpose for which ideally 
the qubit would be in a particular pure state $\ket{\phi})$, as discussed 
in Ref.~\cite{Jac04}. Then the cost function should be  
based on the performance of the qubit, as can be empirically determined 
independent of the preparation procedure (for example, by a second party 
who will impose the cost on the preparer). 
Any such cost function depends 
only upon ${\bf E}\bra{\phi}\rho_T\ket{\phi}$, the expected 
fidelity of the state ${\bf E}[\rho_T]$ with the fiducial state $\ket{\phi}$.
 For example, for $\sigma_z\ket{\phi}=+\ket{\phi}$, the fidelity, which we would 
 wish to maximize, is  $[1+{\bf E}(R_T\cos\theta_T)]/2$. 
 Since we are working under the assumption that 
  $\theta_t$ can be set arbitrarily, all that we require 
 is to maximize ${\bf E}(R_T)$. That is, the best motivated 
 cost function for qubit purification 
 is ${\bf E}_s(-\sqrt{1-S_T})$ rather than ${\bf E}_s(S_T)$. 
 
 We note that Jacobs' argument for the optimality of his control strategy 
 with respect to his cost function relies upon the linearity of that cost function
 in $S$ and hence does not apply to the cost function ${\bf E}_s(-\sqrt{1-S_T})$.
 Nevertheless, below we prove rigorously that Jacobs' strategy is optimal
 for both cost functions, and the calculation is essentially the same in both cases. 
 We do this by considering the general  
 cost function
  \begin{equation}
  J[u,s] = \mathbf{E}_s\big(F(S_T^u)\big),
  \end{equation} 
where $F$ can be given 
either by $F(x) = x$ or by $F(x) = -\sqrt{1-x}$.
Here the superscript $u$ on $S_t^u$ means that 
$S^u_t$ is propagated by \erf{eq system} under 
the control strategy $u$.
   
Before presenting the theorem we use to prove that Jacob's strategy is optimal,
we first give a a heuristic derivation of the so-called 
\emph{Bellman equation}, see e.g.\ \cite{Kus71}. 
This derivation provides some insight into the problem, and the 
Bellman equation will 
reappear in the theorem. In situations (unlike the present case) 
where one did not have a candidate optimal control strategy,
the Bellman equation may also suggest such a candidate. 

To derive the Bellman equation, consider a given Markov 
strategy $u$, and define 
the \emph{cost-to-go} at time $t$ 
given that $S_t =s$ by
  \begin{equation}
  J[u,s,t] = \mathbf{E}\left(F\left(S^u_T\right)\big|S_t =s\right),
  \end{equation} 
so that $J[u,s]=J[u,s,0]$. 
 Now we assume that there exists an 
{\em optimal} control strategy $u_t^\star$; that is, a strategy 
that minimizes the cost function. We also assume that it is \emph{Markov};
that is, $u^\star_t=u^\star(t,S_t)$, where $u^\star(t,s)$ is 
a deterministic function. From these assumptions, we can 
define the \emph{value function} to be the 
{\em optimal cost-to-go}:
  \begin{equation}
  V(t,s) = J[u^\star,s,t].
  \end{equation}

Since $u^\star_t$ is a strategy that minimizes the 
cost over the full interval $[0,T]$ it seems obvious  that it will also 
minimize the cost-to-go over the interval $[t,T]$; that is, 
  \begin{equation}
  V(t,s) = J[u^\star,s,t] = \inf_u J[u,s,t],
  \end{equation}
where the infimum is over all Markov control strategies.  This is a
form of the \emph{the principle of optimality}
and can indeed be made rigorous, see e.g.\ \cite{Kus71}. Since at this point we 
are merely arguing heuristically, we do not need a rigorous proof. 
The idea behind the proof is, however, simple to explain. If a better
strategy $u$ exists on the interval $[t,T]$, we would simply 
use $u^\star$ up to time $t$ and would proceed with the 
better strategy $u$ from time $t$ onwards and obtain 
a lower cost over the interval $[0,T]$. This
contradicts optimality of $u^\star$ and hence such a 
$u$ can not exist. 

The principle of optimality leads to the following recursion for the value 
function (optimal cost-to-go)
  \begin{equation}
  V(t,s) = \inf_u \mathbf{E}\left(V\big(t',S^u_{t'}\big)\big| S^u_t =s\right), \ \ \ \ V(T,s) = F(s),
  \end{equation}
where $0 \le t \le t' \le T$ and the infimum is over 
the Markov control strategies on the interval $[t,t']$.
To see this, note that $\mathbf{E}(V(t',S^u_{t'})| S^u_t =s)$
is the cost-to-go of a control strategy that follows $u$
on the interval $[t,t']$ and $u^\star$ on the interval $[t',T]$.
If we take $t' = t+dt$, expand $V(t',S_{t'})$ according to It\^o's 
rule, and re-arrange the terms, then we arrive at 
the following \emph{Bellman equation}
  \bqa\label{eq Bellman}
  \frac{\partial V}{\partial t}(t,s)  & =& - \inf_{v \in [0,1]} \mathcal{L}^v V(t,s) \\
  &=& 4s\frac{\partial V}{\partial s}(t,s) - \inf_{v\in [0,1]} v^2{\cal D}V(t,s),
\eqa  
with terminal condition $V(T,s) = F(s)$. Since $u^\star$ is an optimal strategy, 
the infimum is attained at $v = u^\star(t,s)$. Moreover, it is clear that $u^\star(t,s) \in \{0,1\}$, 
so that  the Bellman equation can be rewritten as 
\beq \label{HJBE2} 
\frac{\du V}{\du t} -4s\frac{\partial  V}{\partial s} + {\rm min}\cu{ 0, {\cal D} V }= 0.
\eeq

Jacobs' strategy chooses $u_t=0$ for $0\leq t\leq T$. This corresponds to instantaneously rotating the qubit onto the $x$-axis at $t=0$, and then actively maintaining it there. We now show that it yields a 
solution to the Bellman equation, as it should if it were an optimal strategy.  For $u_t=0$ \erf{dS} gives the deterministic solution $S_t = S_0 e^{-4t}$, so that $S_T=S_t e^{-4(T-t)}$. If this strategy were optimal then  $V(t,s)$, the optimal cost-to-go, should be constant at $F(S_T)$, and hence should be given by
 \beq \label{eq VJ}
 V(t,s) = F(s e^{-4(T-t)}).
\eeq
For this choice it is simple to verify that for both 
cases $F(x) =x$ and $F(x) = -\sqrt{1-x}$,  
and for all  $s \in [0,1]$ and $t \in [0,T]$, that 
\bqa
{\cal D}V &\geq & 0,\\
\frac{\du V}{\du t} &=&  4s\frac{\partial  V}{\partial s}.
\eqa
Thus \erf{eq VJ} is indeed a solution of \erf{HJBE2}.

We emphasize that this result does not prove that Jacobs' strategy is optimal. A 
strategy that is optimal will give a solution to the Bellman equation, but a
solution of the Bellman equation does not necessarily correspond to an optimal 
strategy. However, it is simple to check whether it does correspond to an optimal 
strategy, as the theorem below shows. 
Instead of assuming the existence 
of an optimal strategy $u^\star$, we now consider  
 a solution $V$ to the Bellman equation \erf{eq Bellman} 
with terminal condition $V(T,s) = F(s)$ and we 
\emph{define} a candidate optimal Markov strategy
$u^\star$ by
  \begin{equation}\label{eq candidate}
  u^\star(t,s) \in \mathop{\mathrm{argmin}}_{v\in [0,1]}{\mathcal{L}^v V(t,s)},
  \end{equation}
  where the right-hand-side means the 
  value of $v\in [0,1]$ that minimizes ${\mathcal{L}^v V(t,s)}$. 
  
Theorem \ref{thm verijacobs},
which is a special case of a 
standard result in control theory \cite[Thm.\
8.1]{FlS}, provides a set of criteria that guarantees 
that a particular control strategy $u_t$ is optimal.
We apply the theorem to our candidate 
strategy with $V$ as in \erf{eq VJ}.
Note that criteria \ref{two} and \ref{three} together 
are equivalent to $V$ satisfying the Bellman equation and 
$u$ being defined as in \erf{eq candidate}. Criterion \ref{four} 
is the terminal condition for the Bellman equation. 
Criterion \ref{one} is a smoothness condition on 
$V$, which from \erf{eq VJ} is readily seen to hold 
for the two functions $F$ we are considering.
The Theorem therefore allows us to conclude that our 
candidate strategy is indeed optimal.

\begin{theo}{\bf (Verification theorem)} \label{thm verijacobs}
Suppose there exists a Markov control
$u:[0,T]\times [0,1] \to [-1,1]$ and a 
function $V:[0,T]\times [0,1] \to \mathbb{R}$ such that
\begin{enumerate}
\item \label{one} $V(t,s)$ is $C^1$ on $[0,T)$ 
and $C^2$ in $s$.
\item \label{two} For all $t\in[0,T]$ and $s\in[0,1]$,
$$
     \frac{\partial V}{\partial t}(t,s) + \mathcal{L}^{u(t,s)}V(t,s) = 0.
$$
\item \label{three} For all $t\in[0,T]$, $v\in[-1,1]$ and $s\in[0,1]$,
$$
     \frac{\partial V}{\partial t}(t,s) + \mathcal{L}^{v}V(t,s) \ge 0. 
$$
\item \label{four} For all $s\in[0,1]$, $V$ satisfies the terminal condition
$$
    V(T,s) = F(s).
$$
\end{enumerate}
Then the Markov control strategy $u$ is optimal, i.e.\ $J[u,s]\le 
J[\tilde{u},s]$ for all $s\in[0,1]$ and any control $\tilde{u}_t$ taking 
values in $[-1,1]$, and moreover $V(0,s)=J[u,s]$.
\end{theo}
{\it Proof.}
Let $\tilde u_t$ be any control strategy.  Then using Dynkin's formula 
and the first, third and fourth conditions in the verification theorem, we 
obtain
\bqa
	V(0,s) &=& \mathbf{E}_s(V(T,S_T)) \nn \\
	&& - \mathbf{E}_s\int_0^T\!\left[
		\frac{\partial V}{\partial t}(t,S_{t})+
		\mathcal{L}^{\tilde u_{t}}V(t,S_{t})
	\right]\!dt \nn \\
	&\le& \mathbf{E}_s(V(T,S_T)) = \mathbf{E}_s\left(F(S_T)\right) = J[\tilde u,s]. \nn
\eqa
Using the second condition, we similarly conclude that $V(0,s)=J[u,s]$.
Hence $J[u,s]\le J[\tilde u,s]$ for any $\tilde u$.
\hfill$\square$\break

\section{Wiseman and Ralph's problem}
As well as the variation on Jacob's problem discussed above, 
Wiseman and Ralph \cite{WisRal06} also considered a quite different control problem.  
Their problem fixes some threshold $h\in (0,1)$, and seeks a feedback control $u_t$ 
that minimizes the average time at which the linear entropy first 
hits the threshold value of $h$. That is, the cost function is
\begin{equation}
	J[u,s] = \mathbf{E}_s\left(\tau^u\right),\qquad
	\tau^u = \inf\left\{t:S^u_t\le h\right\}.
\end{equation}
As before, the superscript $u$ on $S_t^u$ reminds us 
that $S$ is propagated by \erf{eq system} under 
the control strategy $u$. 
It is clear from the discussion at the start of the preceding section that, 
under our assumptions, this problem is equivalent 
to minimizing the time taken to prepare 
a given pure state (in the $y=0$ plane) with fidelity $(1+\sqrt{1-h})/2$.  

From consideration of the stochastic evolution of 
$\log(1-S_t)$, Wiseman and Ralph \cite{WisRal06} 
claimed that the strategy $u = 1$ was optimal 
in the asymptotic ($h \ll 1$) limit. This strategy 
requires instantaneously rotating (if necessary) the 
qubit onto the $z$-axis at $t=0$, then letting 
$\Omega_t=0$ for $t>0$. Here we verify that this is indeed
the optimal strategy for all $h$. We proceed as 
before, first finding the Bellman equation and  
exhibiting a solution using the strategy 
proposed by Wiseman and Ralph. 
Then we turn the argument around and use 
the solution to the Bellman equation to 
prove rigorously that the candidate strategy is indeed 
optimal. 

To derive the Bellman equation, we again start by assuming the existence of an 
optimal Markov strategy $u^\star_t$. The value function, the  
optimal cost-to-go, is given by
  \begin{equation}
  V(t,s) = \mathbf{E}\left(\big(\tau^{u^\star}-t\big)^+\Big|S^{u^\star}_t =s\right),
  \ \ t\ge 0,\ 1 \ge s \ge h.
  \end{equation}
Here $(\tau^{u^\star}-t)^+ = \max\{0,\, (\tau^{u^\star}-t)\}$, 
the time left after time $t$ until the threshold 
$h$ is first hit. Note that 
it follows immediately from the definition that $V(t,s) =0$ 
for all $t \ge 0$ and $s\le h$. The principle of optimality then yields the 
following recursion for the value function 
(where $t'\wedge\tau^u$ means $\min\{t',\tau^u\}$)
  \begin{equation}\label{eq dynprog}
  V(t,s)  = \inf_{u}\mathbf{E}\left((t'\wedge \tau^u)-t +
  V\big(t',S^u_{t'\wedge\tau^u}\big)\big|S^u_t =s\right).
  \end{equation}  
Here $0 \le t\le t',\ h \le s \le 1$ and the 
infimum is over all Markov control strategies on 
the interval $[t,t']$. To see this, note that 
$\mathbf{E}\left((t'\wedge \tau^u)-t +
  V\big(t',S^u_{t'\wedge\tau^u}\big)\big|S^u_t =s\right)$
is the cost-to-go of a control strategy that follows $u$
on the interval $[t,t']$ and $u^\star$ after time $t'$.
If $t'$ is smaller than $\tau^u$ then we incur a cost $t'-t$ 
on the interval $[t,t']$ plus the optimal cost-to-go 
if we are at $S_{t'}^u$ at time $r$. If $t'$ is greater 
than $\tau^u$ then we incur a cost $\tau^u-t'$ over 
the interval $[t,t']$. Note that $V(t',S_{\tau^u}) = V(t',h)= 0$.  

If we take $t'=t+dt$, expand $V$ according to 
It\^o's rule and re-arrange the terms, then 
\erf{eq dynprog} leads to the following Bellman equation
  \bqa \label{eq BellmanWR} 
  -\frac{\partial V}{\partial t} &=& 1 + \inf_{v\in [0,1]} \mathcal{L}^vV(t,s) \\
  &=& 1 - 4s\frac{\partial V}{\partial s}(t,s) + \inf_{v\in [0,1]} v^2{\cal D}V(t,s),
  \eqa
with boundary condition $V(t,h) = 0$. 
As before, the minimand is linear in $v^2$ so this can be 
rewritten as  
  \beq \label{HJBE3}
  -\frac{\partial V}{\partial t} = 1 
   -4s\frac{\partial  V}{\partial s} + \min\cu{0,{\cal D}V }.
\eeq

Having obtained the Bellman equation, we now need 
to find a solution. Wiseman and Ralph \cite{WisRal06} 
  show by explicit calculation that 
for $u_t = 1$ the expected cost-to-go is
\bqa
   J[u,s,t]&=& \mathbf{E}\left((\tau^u-t)^+\big|S^u_t =s \right) \\
   &=&[\gamma(h) - \gamma(s)]/4 ,
 \eqa
 where for convenience we have defined, for $x \in [0,1]$, 
 \beq \gamma(x) = \sqrt{1-x}\,\mbox{atanh}\big(\sqrt{1-x}\big) .
 \eeq
Note that $J[u,s,t]$ is independent of $t$. This is to be 
expected, since the control strategy $u_t =1$ is 
not time dependent, so the time left until we hit the 
threshold $h$, starting at $s$, does not depend on when 
$s$ was actually reached. Now, if Wiseman and Ralph's 
proposed strategy is optimal, then from the heuristic arguments 
above, taking  
\beq
V(s) = [\gamma(h) - \gamma(s)]/4
\label{eq VWR}
\eeq
should 
satisfy the Bellman equation \erf{HJBE3}.  

It is easily verified that 
for all $s\in [h,1]$  \erf{eq VWR} satisfies
\begin{eqnarray}
4s\frac{\partial V}{\partial s}(s) - {\cal D}V(s) &=& 1,\\
{\cal D}V(s) &\leq& 0,
\end{eqnarray}
as well as $V(h)=0$. Thus the strategy of Wiseman and Ralph does indeed give a solution of the 
Bellman equation \erf{HJBE3}. Moreover, it is clear that the infimum in \erf{HJBE3}
is attained at $v = 1$.  Thus if we turn the problem around and \emph{define} a candidate 
optimal control strategy via 
\erf{eq candidate} using the solution \erf{eq VWR} 
to the Bellman equation, this  gives the Wiseman-Ralph 
strategy $u^\star(t,s) = 1$ for all 
$t\ge 0$ and $s\in [h,1]$. Note that we 
do not need to worry about $s\in[0,h)$---in this case 
$\tau=0$, which can not be improved upon! 

We emphasize again that the above argument only establishes that  
the Wiseman-Ralph strategy is a candidate optimal strategy. To verify that it is  optimal 
we proceed as before, via a second verification 
theorem. This is another special case of a 
standard result in control theory 
\cite[Thm.\ 3.1]{OkS}. We apply the 
Theorem to our candidate strategy with $V$ as in \erf{eq VWR}. 
Note again that criteria \ref{twoa} and \ref{threea} together 
are equivalent to $V$ satisfying the Bellman equation and 
$u$ being defined as in \erf{eq candidate}. Criterion \ref{foura} 
is the terminal condition for the Bellman equation. 
Criterion \ref{onea} is a smoothness condition on 
$V$, which from \erf{eq VWR} can be verified to hold.
This Theorem therefore allows us to conclude that the Wiseman-Ralph strategy is
indeed optimal. 
\begin{theo}{\bf (Verification theorem)}\label{veriwiseman}
Suppose there exist a stationary Markov control 
$u:[h,1] \to [-1,1]$ such that 
$\mathbf{E}_s\left(\tau^u\right) < \infty$, 
and a function $V:[h,1]\to\mathbb{R}$ such that 
\begin{enumerate}
\item V(s) is $C^2$ in $s$. \label{onea}
\item For all $s\in [h,1]$, $\mathcal{L}^{u(s)}V(s) + 1 = 0$. \label{twoa}
\item For all $v\in[-1,1]$ and $s\in[h,1]$, $\mathcal{L}^vV(s) + 1 \ge 0$. \label{threea}
\item V(h) = 0. \label{foura}
\end{enumerate}
Then the stationary Markov control strategy $u$ is optimal, i.e.\ 
$J[u,s]\le J[\tilde{u},s]$ for any control $\tilde u_t$ taking values in 
$[-1,1]$, and moreover $V(s)=J[u,s]$.
\end{theo}

{\it Proof.}  Let $\tilde u_t$ be any control strategy. 
Just as in the proof of the previous theorem, we would  
like to apply Dynkin's formula, but with the {\it random} 
terminal time $\tau^{\tilde u}$ replacing $T$. This is possible only 
if $\mathbf{E}_s\left(\tau^{\tilde u}\right) < \infty$ \cite{Oksendal}. 
However, note that if $\mathbf{E}_s\left(\tau^{\tilde u}\right)$ is 
infinite, then it is immediate that 
  \begin{equation*}
  J[u,s] = \mathbf{E}_s\left(\tau^u\right) < 
  \mathbf{E}_s\left(\tau^{\tilde u}\right) = J[\tilde u, s].
  \end{equation*}
That is, we can safely restrict our attention to control strategies 
$\tilde u$ for which $\mathbf{E}_s\left(\tau^{\tilde u}\right)$ is 
finite. Therefore we can now safely apply Dynkin's formula using 
the random terminal time $\tau^{\tilde u}$ \cite{Oksendal}.  
This gives, using the conditions of the theorem,
\begin{multline*}
	V(s)=\mathbf{E}_s\left[
		V\left(S^{\tilde u}_{\tau^{\tilde u}}\right)-
		\int_0^{\tau^{\tilde u}}\mathcal{L}^{\tilde u_t}
		V\left(S^{\tilde u}_t\right)\,dt
	\right] \\
	\le \mathbf{E}_s\left[
                V\left(S^{\tilde u}_{\tau^{\tilde u}}\right)+\int_0^{\tau^{\tilde u}} dt
        \right]=\mathbf{E}_s\left(V\left(S^{\tilde u}_{\tau^{\tilde u}}\right)+\tau^{\tilde u}\right).
\end{multline*}
But note that $S^{\tilde u}_{\tau^{\tilde u}}=h$ by the definition of $\tau^{\tilde u}$; so
$$
	V(s) \le \mathbf{E}_s\left(V(h)+\tau^{\tilde u}\right) = 
	\mathbf{E}_s\left(\tau^{\tilde u}\right) = J[\tilde u,s].
$$
Using the remaining (second) condition, we similarly conclude that $V(s)=J[u,s]$.  
Hence $J[u,s]\le J[\tilde u,s]$ for any $\tilde u$.
\hfill$\square$\break

\section{Conclusion}
We have highlighted a simple technique from optimal control theory to
verify optimality of candidate control strategies. This paper was
prompted by the absence of such arguments in the current physics
literature (but see \cite{BvH05}). 
In particular, we have verified the optimality of previously 
proposed control strategies for two problems of current interest 
\cite{Jac03,Jac04,WisRal06,ComJac06,Gri07,HilRal07} in 
nonlinear quantum feedback control. 

Optimal control theory is notoriously difficult for nonlinear systems, and
the reader might be surprised at our explicit computations.  This
simplicity is only made possible by the
rather ``violent'' assumption that we can directly control $\cos\theta_t$
rather than $\Omega_t$.  A much more realistic control problem would be
to find an optimal $\Omega_t$ under additional finite energy constraints
on the control.  Such problems, however, are not analytically tractable.
Various strategies with finite controls were investigated numerically for a particular (solid-state) setting in Ref.~\cite{Gri07}, with no claims of optimality.

\acknowledgements
We thank Ramon van Handel for motivating and guiding this work, 
and for many stimulating discussions. HMW acknowledges support by the Australian Research Council and the State of Queensland, and LB acknowledges support by the ARO
under grant number W911NF-06-1-0378.

\end{document}